\documentclass[reprint,superscriptaddress,
 amsmath,amssymb,
 aps,
prl
]{revtex4-2}
\usepackage{color}
\usepackage{graphicx,graphics}
\usepackage[rgb]{xcolor}
\definecolor{darkgreen}{rgb}{0.0,0.7,0.0}

\usepackage{dcolumn}% Align table columns on decimal point
\usepackage{bm}% bold math
\usepackage{hyperref}% add hypertext capabilities
\begin{document}

\preprint{APS/123-QED}
%From non-icosahedral to icosahedral structures in protonated argon clusters

%Protnated argon culster: icosahedral structure is not always the solution

% Proton coordination-induced structural transtions in charged argon clusters

% Proton coordiantion inducing structrucural transformation in argon clusters

% Unvailing the geometry of protonated argon clusters

%Quantal structural transitions in protonated argon clusters

%How many-body interactions explain structural transtions in protonated argon clusters

% Magic numbers of protonated argon clusters: many-body interactions are essential

% The necessecity of including many-body interactions to describe structural transitions in protonated argon clusters

%Structural transitions in protonated argon clusters: on the role of many-body interactions 

%From pentagonal proton coordination to icosahedral argon clusters

\title{Coordination-driven magic numbers in protonated argon clusters}%
\author{Saajid Chowdhury}
\affiliation{Department of Physics and Astronomy, Stony Brook University, Stony Brook, New York, NY 11794, USA}%
\author{Mar\'{\i}a Judit Montes de Oca-Estévez}
\affiliation{Instituto de Física Fundamental, IFF-CSIC, Serrano 123, Madrid 28006, Spain}%

\author{Florian Foitzik}
\affiliation{Institut für Ionenphysik und Angewandte Physik, Universität Innsbruck, Technikerstraße 25, Innsbruck 6020, Austria}%
\author{Elisabeth Gruber}
\affiliation{Institut für Ionenphysik und Angewandte Physik, Universität Innsbruck, Technikerstraße 25, Innsbruck 6020, Austria}%
\author{Paul Scheier}
\affiliation{Institut für Ionenphysik und Angewandte Physik, Universität Innsbruck, Technikerstraße 25, Innsbruck 6020, Austria}%

\author{Pablo Villarreal}
\affiliation{Instituto de Física Fundamental, IFF-CSIC, Serrano 123, Madrid 28006, Spain}%

\author{Rita Prosmiti}
\affiliation{Instituto de Física Fundamental, IFF-CSIC, Serrano 123, Madrid 28006, Spain}%

\author{Tom\'as Gonz\'alez-Lezana}
\affiliation{Instituto de Física Fundamental, IFF-CSIC, Serrano 123, Madrid 28006, Spain}%

\author{Jes\'us P\'erez-R\'ios}
\affiliation{Department of Physics and Astronomy, Stony Brook University, Stony Brook, New York, NY 11794, USA}

%\homepage{http://www.Second.institution.edu/~Charlie.Author}

\date{\today}% It is always \today, today,
             %  but any date may be explicitly specified

\begin{abstract}
The structural properties of rare-gas clusters can be primarily described by a simple sphere packing model or by pairwise interactions. Remarkably, adding a single proton yields a large set of magic numbers that has remained unexplained. In this Letter, we unravel their origin by combining quantum Monte Carlo techniques with many-body ab initio potentials that correctly capture the proton's coordination environment. Thanks to this approach, we find that argon atoms are mainly localized around the classical minimum, resulting in a particularly rigid behavior in stark contrast to lighter rare-gas clusters. Moreover, as cluster size increases, we identify a clear structural transition from  many-body coordination-driven stability to a regime dominated by two-body interactions, reflecting a reshaping of the underlying potential energy landscape.

%Furthermore, as the magic number of protonated argon clusters increases, a structural transition occurs, reflecting a shift from many-body to two-body interactions in the potential energy landscape. 
 
\end{abstract}

%\keywords{Suggested keywords}%Use showkeys class option if keyword
                              %display desired
\maketitle

%\tableofcontents

%%%%% Intro%%%%%%%%%%%%%

Ion solvation, one of the fundamental processes in chemical physics, is the mechanism by which neutral solvent molecules surround a charged solute as a consequence of solvent-solute interactions. As a result, the solute usually becomes embedded in solvent molecules, yielding a particular shell structure--the solvation shell, whose properties depend on the solute as well as on the solvent. Ion solvation is essential in a wide range of phenomena from energy conversion~\cite{Beatriz} to cold molecular sciences~\cite{Saajid2024,astrakharchik2020ionic,Bruun}. Most of the literature on solvation follows a macroscopic approach using thermodynamics and chemical kinetics. However, it is possible and preferable to explain solvation at the microscopic level, where only a full quantal treatment is meaningful~\cite{QMC1,QMC2,QMC3,GarciaAlfonso22}, although, in principle, in large systems, only a classical approach is considered practical~\cite{Classical1,Classical2}. Along these lines, there is extensive literature on solvation, particularly on ion solvation~\cite{Bartolomei2021,TomasReview,Josu2019,Zunzunegui2023,Roszak2003,Rastogi2018}, focusing on the stability (structure) of different cations and anions in the presence of a quantum solvent--atomic or molecular. Thus, recent investigations of doped helium clusters have focused on understanding either the first steps of the solvation of the ion \cite{Albrechtsen2023, ACPSCSBGPS:JCP2025, GBHP:JCP24, Calvo24,D5CP00318K,DEOCAESTEVEZ2025142242} or formation of large well-ordered solvation shells around the ion \cite{Zunzunegui2023,di-cation}. Experimentally, the stability of different ion-solvent clusters is readily accessible, yielding the so-called magic numbers, defined as the numbers of solvent atoms or molecules in the most stable clusters with the highest evaporation energies.

%Experimentally, the stability of different ion-solvent clusters is readily accessible, yielding the so-called magic numbers, defined as the number of solvent atoms or molecules in the most stable clusters with the highest evaporation energies. 

%that maximize the evaporation energy of the cluster

Most experimental data on ion solvation show a limited set of magic numbers, even in the case of di-cations~\cite{Zunzunegui2023,di-cation}. Most of these studies show that quantum Monte Carlo techniques can sufficiently replicate the observed sequence of magic numbers and explain the nature of the clusters. One exception, though, is the proton solvation in Ar clusters, showing a large set of magic numbers~\cite{Gatchell2018,Gatchell2019}: 7, 13, 19, 23, 26, 29, 32, 34, 46, 49, and 55, that served to explain the experimental findings of Harris et al. on the study of charged argon clusters~\cite{HARRIS1986316,Harris}. Despite all theoretical efforts, only some of these--7, 13, and 19, have been demonstrated in quantum chemistry calculations~\cite{Gatchell2018,McDonald2016,Montes2024,Montes2024-cpl,Giju2002,molecules29174084,Montes2021}. The rest have been qualitatively described resorting to sphere packing arguments in light of the similarity between the set of magic numbers of protonated argon and Lennard-Jones clusters~\cite{Gatchell2019,Gatchell2018}. 
 However, these results neglect the chemical properties of the ion encoded in its coordination number, i.e., the number of bonds it forms. In other words, most approaches to date have not considered the ion's binding mechanism. Indeed, it has been suggested that the coordination number of the ion may have a drastic influence on the structure and magic number sequence in ion solvation~\cite{Roszak2003}.

%and  qualitatively described, have five-member ring filling structures around the charge, but for charged Ar clusters rather than for protonated Ar clusters, or assuming icosahedrall shell completion~\cite{Harris,HARRIS1986316}. In addition, although sometimes forgotten, the coordination number, i.e., the number of bonds a chemical forms, may have a drastic influence on the structure and magic number sequence of an ionic cluster~\cite{Roszak2003}. 

In this Letter, we present a combined experimental and theoretical investigation of protonated argon clusters. On the one hand, new experimental results on the system have been performed as an extension of the previously reported study by Gatchell {\it et al.} \cite{Gatchell2018,Gatchell2019}. On the other hand, we present results from a full quantum-mechanical analysis of the stability of these clusters using quantum Monte Carlo methods, based on an accurate {\it ab initio} potential energy surface (PES) that describes the proton's coordination, explaining all the experimentally observed magic numbers. We find that the proton is not always at the center of the most stable clusters. Moreover, we show that the rule-of-thumb for magic numbers based on layer filling around the charge is inadequate. Instead, we find a structural transition as the magic number series progresses, from a pentagonal-ring filling geometry to an icosahedral geometry that pushes the proton to one side of the cluster's central axis, related to the role of ligand-ligand versus ion-ligand interactions. This transition expresses a more profound change in the relevance of few-body versus many-body interactions. 

\begin{figure*}
    \includegraphics[width=1\textwidth]{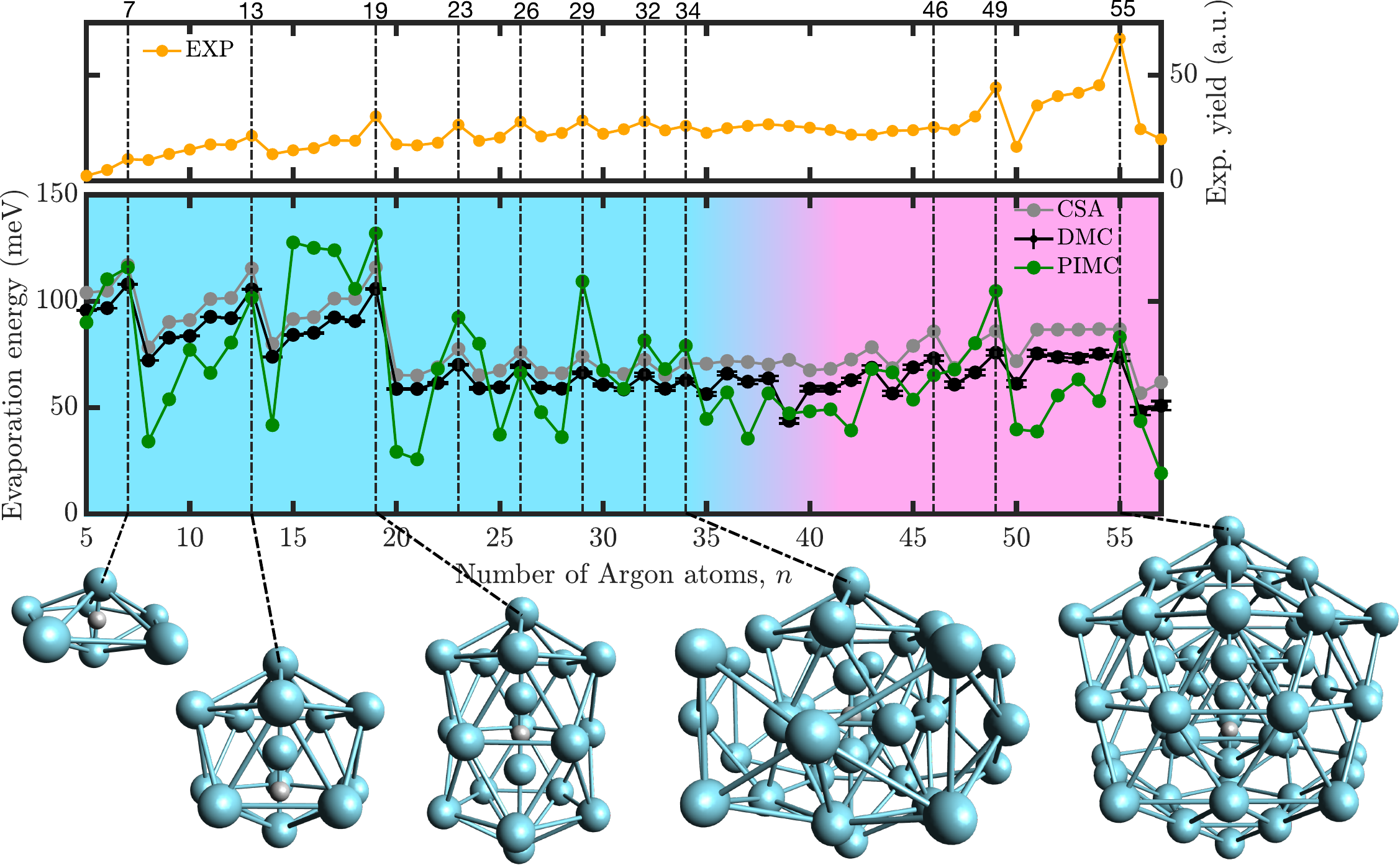}
    \caption{(Top panel) Experimentally observed ion abundances as a function of the number of argon atoms. Due to the top-down fragmentation process, these abundances reflect the relative stability of the clusters. (Bottom panel) Absolute value of theoretical evaporation energies of Ar$_n$H$^+$ clusters. The gray points show the potential energy differences $V_n-V_{n-1}$ computed using the classical optimization technique CSA. The black points show the ground state energy differences $E_n-E_{n-1}$, accounting for the zero-point energy in the vibrational motion of the nuclei, computed using DMC. The green points show the energy differences $\overline{E}_n-\overline{E}_{n-1}$, which also accounts for the thermal distribution, computed using PIMC. The dotted lines show the experimentally-observed magic numbers of Ar$_n$H$^+$ clusters. The clusters Ar$_n$H$^+$ in the blue region have a second shell consisting of pentagonal rings, while the clusters in the magenta region have an icosahedral second shell. This structural change correlates with a transition from many-body effects (blue region) to few-body effects (magenta). The classical structures are shown for $n=7$, 13, 19, 34 and 55; from left to right in the plot.}
    \label{fig1}
\end{figure*}

The new experimental results presented in this Letter were obtained using an experimental setup described in detail elsewhere~\cite{bergmeister2023}. In brief, protonated argon clusters were formed inside helium nanodroplets. Pre-ionized, multiply charged helium droplets~\cite{laimer2019} were first doped with molecular hydrogen gas, which is attracted to the charges across the droplet surface and ionized via charge transfer to form (H$_2$)$_n$H$^+$. Subsequent doping with argon induces proton transfer to the first argon atom that encounters each (H$_2$)$_n$H$^+$ ion~\cite{tiefenthaler2020}. Additional argon atoms then solvate these ArH$^+$ ions, forming Ar$_n$H$^+$ cluster ions. 

A key advantage of the new setup is that it enables precise, controlled fragmentation of these clusters. To release the Ar$_n$H$^+$ complexes from the droplets, the droplets collide with room-temperature helium gas. This controlled collision reduces the droplet size, and single helium-solvated Ar$_n$H$^+$ clusters are ejected. By adjusting the helium-evaporation pressure to be sufficiently high, we remove not only the remaining helium atoms but also weakly bound argon atoms, ensuring that only the most stable cluster configurations persist. This top-down fragmentation directly links the observed ion abundances to evaporation energies and, thus, to cluster stability; higher abundances indicate greater stability~\cite{hansen1999}. The resulting argon series are measured via time-of-flight mass-spectrometry and presented in Figure~\ref{fig1}.

%%%%%%%%%%%%%%%%%%%%%%%%%%%%

%Implication for seeded nucleation. In nucleation it is commong to use pair-wise interaction, so it is only valid in the case of large clusters not for any cluster size. This is a very important message!

%55B Mackay icosahedron

%Describe potential energy surface

In all calculations reported here, we employ a many-body expansion formalism~\cite{Murrell1984} to model the underlying interactions as the number of Ar atoms in the Ar$_n$H$^+$ increases~\cite{molecules29174084}, 

%a Lennard-Jones potential to characterize the Ar-Ar interaction, whereas the proton-Ar interaction is described by a many-body expansion formalism~\cite{Murrell1984} to model the underlying interactions as the number of Ar atoms increases in the clusters as~\cite{molecules29174084},
\begin{equation}
\label{fullPES}
V \approx \sum_i V_i^{(1\mathrm{B})} + \sum_{i<j} V_{ij}^{(2\mathrm{B})} + \sum_{i<j<k} V_{ijk}^{(3\mathrm{B})} + \sum_{i<j<k<l} V_{ijkl}^{(4\mathrm{B})},
\end{equation}
where the two-body (2B) terms ($V_{\mathrm{ArH}^+}$, $V_{\mathrm{Ar}_2}$), the three-body (3B) term ($V_{\mathrm{Ar}_2\mathrm{H}^+}$), and the four-body (4B) term ($V_{\mathrm{Ar}_3\mathrm{H}^+}$) are represented by machine-learned potential energy surfaces trained on gold-standard CCSD(T)/CBS and CCSD(T)-F12/aug-cc-pVQZ data~\cite{Montes2021,Montes2024,Montes2024-cpl}. This expansion framework enables an accurate and transferable description of cooperative effects in the PES, explicitly incorporating the proton’s coordination number of 2, and thereby accounting for the 4B [Ar–H–Ar]$^+$–Ar interactions.

The stability of protonated argon clusters is studied at three different levels of theory. First, we use the conformational space annealing (CSA) method \cite{Lee2003} to calculate the global minima of Ar$_n$H$^+$, which is a classical optimization technique. The algorithm starts with a population of randomly generated configurations. These configurations undergo mating, mutation, and gradient descent, in order to evolve the population. To maintain the diversity of local minima, a structural dissimilarity metric is used. The algorithm was run at least 25 times for each $n$, with up to two million local minimizations for each run. On the other hand, we calculate the ground state energy of every cluster using the Diffusion Monte Carlo (DMC) method \cite{Kosztin1995}, including the zero-point energy correction due to the vibration of the atoms around the global minimum. For each $n$, we run at least 25 independent runs of DMC, where in each run, 20,000 walkers initially start at the configuration given by CSA. For 10,000 timesteps ($d\tau=10$ a.u.), the walkers perform Gaussian random steps, including the rotation and translation of the [Ar–H–Ar]$^+$ rigid core. Depending on the difference in potential energies before and after the timestep, the walker is either killed or replicated up to 2 times. The average potential energy of the walkers during the final 2500 timesteps is reported as the ground state energy. Ultimately, to explore the stability of a given cluster is to include temperature effects; in our case, we employ a path integral Monte Carlo (PIMC) method \cite{QMC1} to obtain both energies and geometrical structures of the corresponding Ar$_n$H$^+$ clusters at $T = 2$ K. Simulations for 10,000 steps are also initiated from classically optimized configurations considering between 400 and 500 of the so called quantum beads $M$, using a thermodynamic energy estimator in the same way as recent investigations of ion-doped helium clusters~\cite{Zunzunegui2023, di-cation, ACPSCSBGPS:JCP2025}.

The theoretical predictions for the evaporation energies as a function of the number of atoms within a cluster are shown in Fig.~\ref{fig1} against the experimental data (upper panel), where vertical dashed lines identify the magic numbers. The classical results are displayed in gray, the DMC ones in black, and the PIMC in green. Remarkably, the classical results describe all the magic numbers. The classical and DMC evaporation energies are very close to each other, indicating a rigid structure for protonated argon clusters. This is ultimately confirmed by the PIMC, showing minor deviations from the DMC results. The first magic number $n=7$ (pentagonal bipyramid) is the result of a pentagonal ring around the axis of the coordination complex [Ar-H-Ar]$^+$. The following magic number, $n=13$, is the first icosahedron, as observed for Lennard-Jones clusters. Notably, the ion is not at the center of this cluster. However, for the following magic number $n=19$, as a result of adding two pentagonal pyramids (at the top and the bottom) of the pentagonal bipyramid exhibited by Ar$_{7}$H$^+$, the proton is at the center of the cluster. These three magic numbers have previously been characterized using quantum chemical calculations and classical minimization methods~\cite{molecules29174084,Gatchell2018}. Next, a set of several numbers with a difference of 3 and 2 argon atoms appears (23, 26, 29, 32, and 34) as more argons describe a new shell around the proton, until they reach $n=39$, where a structurally-different half-icosahedral shell is complete (see Fig.4 of the Supplementary Information (SI)). The final magic numbers $n=$~46, 49, and 55 appear as the full icosahedral shell is complete at $n=55$ around the $ n=13$ icosahedron. Such a particular set of magic numbers is the result of the inclusion of many-body interactions, up to 4, in Eq.~(\ref{fullPES}).

%for the particular coordination number of the proton

%very specific set of magic numbers of protonated argon cluster must depend on the itneraction potential beyond the standard two-body interactions, as shown below. 

%The reason for this intriguing behavior is the nature of the coordination complex that serves as the seed for all the clusters. The characterization of the coordination complex is possible thanks to using a many-body interaction potential. 

%The reason behind this is a structural change in the filling of shells, going from a pentagonal ring filling structure around the axis of the complex to an icosahedron-based filling, as usual in rare-gas clusters~\footnote{Indeed, the n=55 corresponds to an icosahedron, as shown in Fig.~\ref{fig1}.}. These observations indicates the relevance of the nature of the coordination complex captured by the many-body nature of the potential energy surface employed in our quantum Monte Carlo simulations. 

\begin{figure}
    \includegraphics[width=0.5\textwidth]{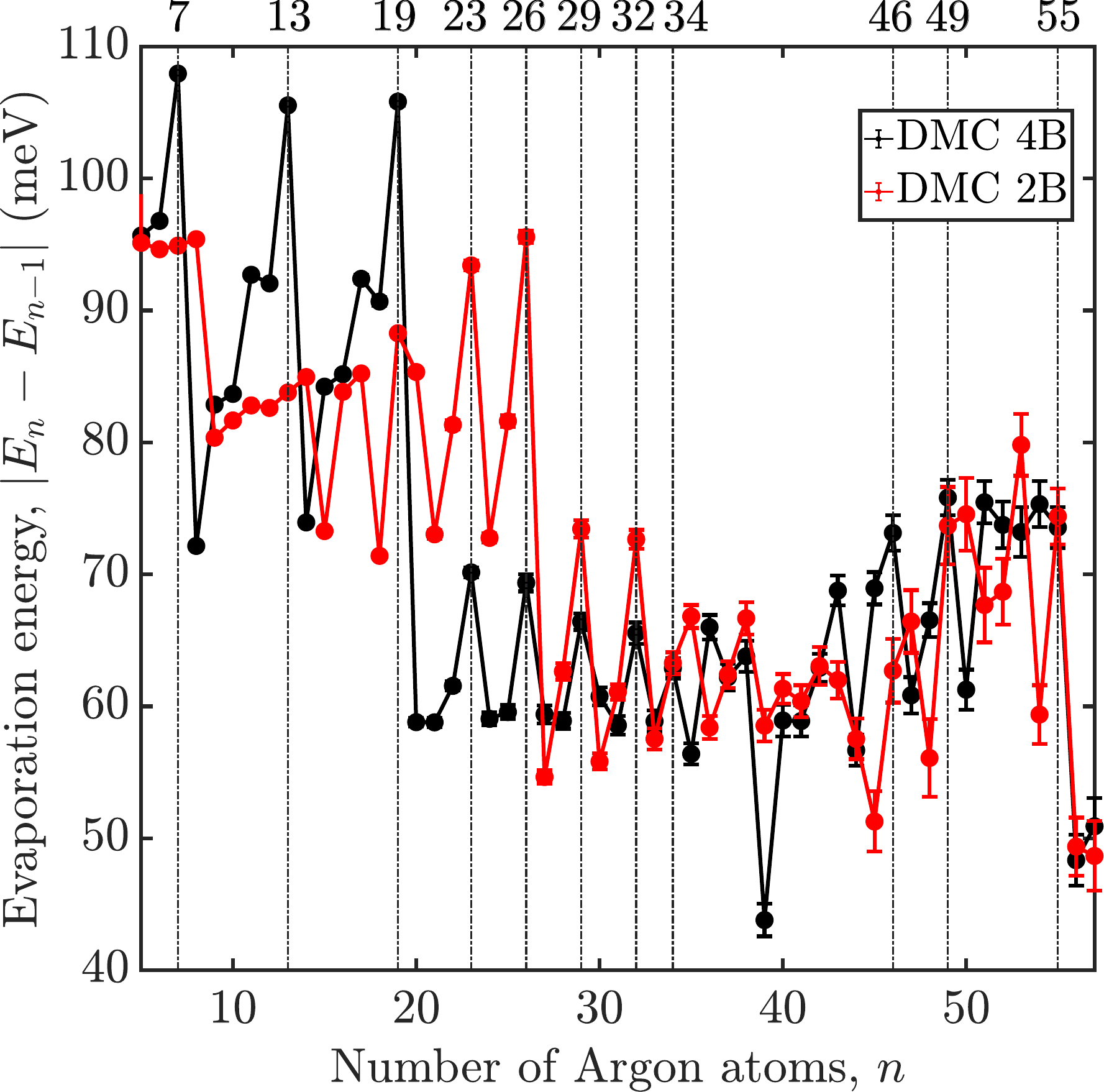}
    \caption{Theoretical evaporation energies of Ar$_n$H$^+$ clusters. The black points show the ground state energy differences $E_n-E_{n-1}$, accounting for the zero-point energy in the vibrational motion of the nuclei, computed using DMC, including the 3B and 4B terms. The red points are for a DMC calculation including only pair-wise interaction potentials.}
    \label{fig2}
\end{figure}

The relevance of these many-body terms, up to 4B, manifests itself in the comparison shown in Figure \ref{fig2}, where the results obtained by using only a pairwise description of the interaction potentials (Ar-Ar and Ar-H$^+$ interactions taken from \cite{Brown2020}~\footnote{The magic numbers and cluster structure are almost identical when using the interaction potential from Ref.~\cite{Tong_2010}.}) are compared with those obtained with the potential~\cite{molecules29174084}, as described in Eq.~(\ref{fullPES}). The evaporation energies in that figure have been obtained exclusively with the DMC calculation, but the corresponding figure with the PIMC is equally illustrative for showing the distinct performance of these two PES. The most striking difference is the peak for the smallest magic numbers. Thus, the features at $n = 7, 13$, and 19 are completely missing when the simulations are performed exclusively with the 2B potential. This anomaly leads us to distinguish between a many-body physics regime (up to $n\approx 34$-39) and a few-body physics regime ($n\gtrsim 39$) in protonated Ar clusters. The present results thus show that the maxima observed in the experimental ion yields for these three Ar$_7$H$^+$, Ar$_{13}$H$^+$ and Ar$_{19}$H$^+$ clusters are only reproduced as anomalous features in the theoretically obtained evaporation energies when these are calculated using a PES that explicitly includes up to 4B effects as discussed in Eq.~(\ref{fullPES}). Indeed, those three clusters are a direct consequence of the structure and properties of the [Ar-H-Ar]$^+$ coordination complex. Then, the configurations responsible for the peaks observed for $n = 23, 26, 29, 32,$ and finally 34 are built by adding Ar atoms, leading to the formation of an equatorial shield of more pentagonal pyramids. The same applies to larger magic numbers up to $n=55$, completing the icosahedron shell.

%The analysis of these three structures reveals that the overall geometry of Ar$_{19}$H$^+$ is the result of adding two pentagonal pyramids (at the top and the bottom) of the pentagonal bipyramid exhibited by Ar$_{7}$H$^+$.

%taken from Ref. \cite{Tong_2010} for Ar-Ar and from 
%Ref. \cite{Brown2020} for Ar-H$^+$

\begin{figure}
    \includegraphics[width=0.45\textwidth]{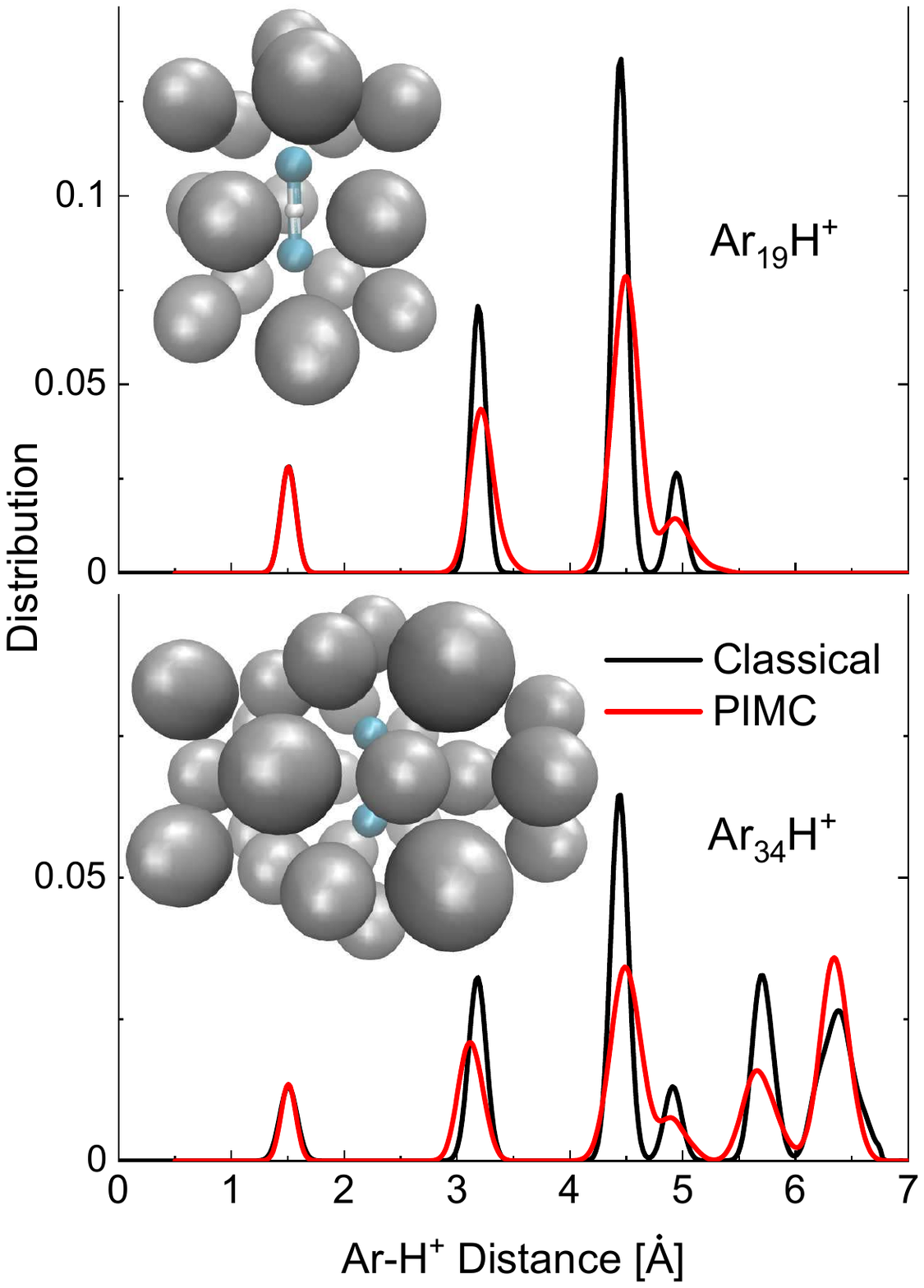}
    \caption{Ar-H$^+$ distance distributions obtained 
    from the classical configurations used to initiate the QM 
    simulations (black) and by means of the PIMC
    calculation (red) for the Ar$_{19}$H$^+$ (top) and Ar$_{34}$H$^+$ (bottom) clusters.
    The insets show probability distributions for both clusters.}
    \label{fig3}
\end{figure}

PIMC simulations, on the other hand, indicate that Ar clusters do not move significantly from the positions occupied in the classically optimized configurations used to initiate the quantum calculations. Ar-H$^+$ distance distributions for the $n=19$ and $n=34$ clusters obtained from the CSA configurations and from the PIMC calculation are plotted in Figure~\ref{fig3}. The comparison between the two distributions reveals that the quantum results do not extend much beyond the peaks in the classical distribution. Figure~\ref{fig3} also includes insets showing the PIMC distributions for both clusters. Additional illustration of the movements of the Ar atoms during the simulation are shown in the animation given in the SI for $n=34$. Larger clusters are thus constructed after a noticeable structural change experienced at Ar$_{39}$H$^+$. Up to this particular size, the cluster structures comprise an inner core formed with the above discussed geometry observed for Ar$_{19}$H$^+$, surrounded by three outer equatorial pentagonal rings of atoms. But, precisely for Ar$_{39}$H$^+$, one of the inner core pentagons (originally antiparallel to the pentagon containing the proton at its center), shifts into a parallel position. Beyond this size, this new arrangement is the foundation which finally yields the large icosahedral symmetrically ordered structure seen for Ar$_{55}$H$^+$. In Figs. 3-5 of the SI, we show the geometrical structures for the Ar$_{19}$H$^+$, Ar$_{39}$H$^+$, and Ar$_{55}$H$^+$ clusters, highlighting this change in the Ar atom positions. The analysis of all these configurations reveals that the construction of the Ar solvation layer around the H$^+$ is carried out in such a manner that, for some sizes, the inner proton does not necessarily occupy the central position, as opposed to some other doped He clusters \cite{di-cation,Zunzunegui2023}.

In summary, by employing a many-body expansion of the interaction potentials, including terms up to 4B order, which explicitly accounts for the proton’s coordination number, we rationalize the full sequence of magic numbers for clusters containing a single proton solvated in argon. The progression of magic numbers arises from a structural transition in Ar$_n$H$^+$, driven by the changing relative importance of many-body versus two-body interactions. The proton–argon coordination complex involving two Ar atoms is crucial to understanding the structure and stability of all magic clusters up to $n=34$, and it seeds the growth of the larger icosahedral motifs observed at larger sizes. However, beyond $n=34$ the coordination complex ceases to play a dominant role with the Ar–Ar interactions taking over, and an effective two-body potential can accurately describe the resulting icosahedral structures. Furthermore, our quantum calculations show only negligible motion of the Ar atoms around their classical positions, indicating a high degree of rigidity across all clusters. These findings have direct implications for impurity physics, particularly for ionic impurities in quantum degenerate gases, where many-body bound states and polarons are expected to form~\cite{astrakharchik2020ionic,Bruun,Meso}. In such systems, the role of coordination complexes has only recently begun to be explored, and so far predominantly from a two-body perspective~\cite{Saajid2024}. In a similar vein, a single ion in a polar gas offers new possibilities to study exotic quantum matter~\cite{Leon}. However, this scenario still lacks a proper theoretical study of the stability of ion-molecule complexes in the ultracold regime, where coordination and the chemical properties of the ion impurity could be essential. Our results, therefore, suggest that explicitly incorporating coordination complexes can significantly affect the predicted structures and dynamics of charged impurities in ultracold gases.

 %In summary, thanks to using a many-body interaction potential accounting for the coordination number of the proton, we have explained all magic numbers of clusters of a single proton solvated in argon. The progression of magic numbers is the consequence of a structural change in protonated argon clusters due to the changing relative importance of many-body interactions compared to two-body interactions. The nature of the coordination complex of the proton with two argon atoms is essential to understand the structure for all clusters with magic numbers up to 34. Beyond this size, the coordination complex does not play a significant role, and the larger icosahedral structures are accordingly explained adequately by a two-body potential. Furthermore, the quantal calculations reveal negligible motion of the argon atoms around their classical positions, which implies a high rigidity for all the clusters. Our findings have implications in impurity physics, especially in the case of an ionic impurity in a quantum degenerate gas, where many-body bound states and polarons are expected to exist~\cite{astrakharchik2020ionic,Bruun,Meso}. Indeed, in those scenarios, only recently has the relevance of the coordination complex been explored, but only from a two-body perspective~\cite{Saajid2024}. Therefore, it is expected, based on our results, that the inclusion of coordination complexes may have an effect on the dynamics of charged impurities in ultracold gases. 

S. C. and J. P.-R. acknowledge the support of the United States Air Force Office of Scientific Research [grant number FA9550-23-1-0202]. M.J.M.O.E., R.P., P.V. and T.G.L. acknowledge funding from Grant No. MICIU/AIE/10.13039/501100011033 PID2024-155666NB-I00 and Comunidad de Madrid grant INVESTIGO ref: INV23-IFF-M2-09. This research was funded in part by the Austrian Science Fund (FWF) [10.55776/I6221, 10.55776/P34563, 10.55776/V1035].

The data and code for CSA and DMC and the cluster geometries are available at \cite{github}.

\end{document}